\documentclass[twocolumn,superscriptaddress,amsmath,amssymb,aps]{revtex4}

\usepackage{graphicx}
\usepackage{dcolumn}
\usepackage{bm}

\newcommand{\s}{\text{\scriptsize sys}}

\newcommand{\T}{{\rm total}}
\newcommand{\dg}{\dagger}

\newcommand{\nl}{\nonumber \\}
\newcommand{\la}{\langle}
\newcommand{\ra}{\rangle}

\newcommand{\ep}{\epsilon}
\newcommand{\w}{\omega}

\newcommand{\be}{\begin{equation}}
\newcommand{\ee}{\end{equation}}
\newcommand{\bea}{\begin{eqnarray}}
\newcommand{\eea}{\end{eqnarray}}
\newcommand{\bsube}{\begin{subequations}}
\newcommand{\esube}{\end{subequations}}
\newcommand{\Eq}[1]{Eq.\,\eqref{#1}}

\newcommand{\Fig}[1]{Fig.\,\ref{#1}}

\newcommand{\comments}[1]{}

\begin{document}

\title{Non-perturbative Theory of Pauli Spin Blockade}

\author{WenJie Hou}
 \affiliation{Department of Physics, Renmin University of China, Beijing 100872, China}
\author{Dong Hou}
 \affiliation{Hefei national laboratory for physical sciences at the microscale, University of science and technology of China, Hefei, Anhui 230026, China}
\author{YuanDong Wang}
 \affiliation{Department of Physics, Renmin University of China, Beijing 100872, China}
\author{JianHua Wei}\email{wjh@ruc.edu.cn}
 \affiliation{Department of Physics, Renmin University of China, Beijing 100872, China}
\author{YiJing Yan}
 \affiliation{Hefei national laboratory for physical sciences at the microscale, University of science and technology of China, Hefei, Anhui 230026, China}

\begin{abstract}
Pauli spin blockade (PSB) is a significant physical effect in double quantum dot (DQD) systems. In this paper, we start from the fundamental quantum model of the DQD with the electron-electron interaction being considered, and then systematically study the PSB effect in DQD by using  a recently developed non-perturbative method,  the hierarchical equations of motion (HEOM) approach. The physical picture of the PSB is elucidated explicitly and  the gate voltage manipulation is described minutely, which are both qualitatively consistent with the experimental measurements. When dot-dot exchange interaction is involved, the PSB effect may be lifted by the strong antiferromagnetic exchange coupling.

\end{abstract}

\maketitle

\section{\label{Intro}INTRODUCTION}
Pauli spin blockade (PSB) is an important physical effect appears in double quantum dot (DQD) systems, which was discovered experimentally in vertically coupled GaAs/AlGaAs DQD as early as 2002 \cite{ono2002vertical}. The basic picture is that the hopping of electrons between two dots will be influenced by their spin configuration if the total excess electron number of the system is 2 with occupation state $(N_1, N_2)=(2,0)$, (1,1) or (0,2), as a consequence, the current - voltage($I-V$) curve will show a rectification behaviour. Obviously, the PSB is caused by the universal Pauli's exclusion principle. It receives extensive studies in various quantum dot systems with different structures from vertical to lateral dots \cite{johnson2005lateral} and from double to three dots \cite{busl2013bipolar}, as well as in dots with different semiconductor materials from GaAs/AlGaAs to Si \cite{liu2008silicon}. Recently, the PSB has been used to fabricate and readout the singlet-triplet spin qubit, which will promote the development of the quantum information \cite{hao2014electron}.

Some important characters  in the PSB regime have been investigated by virous theoretical groups.  Those include: 1) the correlation between the PSB effect and occupation of the two-electron triplet state \cite{fransson2006pauli}; 2) the dynamical nuclear spin polarization by hyperfine interaction \cite{deng2005nuspin}; 3) the nonthermal broadening effect of tunneling current \cite{Kuo2011raeffect}; 4) the leakage-current line shapes from inelastic cotunneling  \cite{Coish2011cotunneling};  5) the spin-flip phonon-mediated charge relaxation in double quantum dots \cite{Danon2013crelax}; and 6) the PSB and the ultrasmall magnetic field effect in organic magnetoresistance \cite{Danon2013UMag}. The Pauli master equation (PME) with second order Fermi¡¯s golden rule is the main approach in above works to archive the transition rates. Other approaches (such as nonequilibrium Green¡¯s functions) are not yet so popular in literatures \cite{Stepanenko2012stsplit,Hsieh2012eprop,Amaha2014trielec}.

We would like to comment that the PME is not accurate enough for the PSB theory. Firstly, the DQD is a typical quantum open system  with infinity degree of freedoms of the total density matrix, while the PME only concerns the diagonal terms of the reduced density matrix and treats the dot-electrode couplings by low-order perturbation schemes; and secondly, the DQD is also a typical strongly correlated system with infinity degree of freedoms of the electron-electron ({\em e-e}) interactions, while the PME either neglects this important interaction or treats it in the single electron level.

Obviously, for the theoretical study on such fundamental physics processes as the PSB,  a non-perturbative approach is highly required to deal with the basic quantum model involving the {\em e-e} interactions.  The  hierarchical equations of motion (HEOM) approach  we newly developed can meet this requirement, which  nonperturbatively resolves the combined effects of dot-electrode dissipation, {\em e-e} interactions, and non-Markovian memory \cite{Jin08234703,Li2012dresponse,2008jcp184112,2009jcp124508,2008njp093016,2009jcp164708,Zhe121129,2015Cheng033009}. In this paper, we start from the Anderson multiple impurity model to describe the DQD, fully considering the {\em e-e} interaction and the dot-electrode couplings. By using the HEOM approach, we deal with this quantum model non-perturbatively to accurately obtain some observations, such as the spectral function, occupancy of electron spin and current, etc. Our theory not only can reveal the physical picture of the PSB clearly but also can elucidate its dependence on various parameters such as  the gate voltages, dot-dot coupling, and exchange correlation between spins in different dots. Besides, the external field manipulation is also convenient to be involved in our theory.

The paper is organized as follows. In Sec.II we briefly review our model and the non-perturbative HEOM approach.   In Sec.III,
we present our accurate solutions relating to the PSB effect, those results include: III.A. the physical picture of the PSB; III.B. the gate voltage manipulation of the PSB; and III.C. the lift of the PSB by the dot-dot exchange interaction.  In Sec.IV we give the summary of our work.

\section{\label{Theo}THEORY AND FORMULAS}

The HEOM is a general formula for the quantum open systems composed of three parts: the system (quantum dots here), the bath (two electrodes here) and the system-bath couplings.  Let us introduce the total Hamiltonian (Anderson impurity model) for the DQD as follows,
\begin{align}\label{ha}
   H_{T}=H_{S}+H_{B}+H_{SB}
\end{align}
where $H_{S}$ is the Hamiltonian for the two coupled dots
 \begin{align}\label{hs}
      H_{S}=\sum_{i=1,2\sigma} \epsilon_{i\sigma}\hat{a}^\dag_{i\sigma}\hat{a}_{i\sigma} + \frac{U}{2}\sum_{i=1,2\sigma} n_{i\sigma}n_{i\bar{\sigma}}
       \nl
      +t\sum_{\sigma}(\hat{a}^\dag_{1\sigma}\hat{a}_{2\sigma}+\text{H.c.})
 \end{align}
here $\epsilon_{i\sigma}$ indicates the on-site energy of the electron with spin $\sigma$ ($\sigma=\uparrow,\downarrow$) on dot $i (i=1,2)$,  $\hat{a}_{i\sigma}^\dag$ and $\hat{a}_{i\sigma}$
correspond the creation and annihilation operators for an electron with spin $\sigma$. $n_{i\sigma}=\hat{a}^\dag_{i\sigma}\hat{a}_{i\sigma}$ is the electron number operator of dot $i$, and $U$ is the Coulomb
interaction between electrons with spin $\sigma$ and $\bar{\sigma}$ (opposite spin of $\sigma$) within one dot. $t$ is the inter-dot coupling, determined by the overlapping integral of electron wave functions. $\text{H.c.}$ stands for the Hermitian conjugate.

For brevity, in what follows, we use the symbol $\mu$ to denote the electron orbital (including spin, space, \emph{etc.}) in the system , i.e.,  $\mu=\{{\sigma},i...\}$. The Hamiltonian of the electrodes is described as a noninteracting Fermi bath,
  \begin{align}\label{hb}
  H_{B}=\sum_{k\mu\alpha=L,R}\epsilon_{k\alpha}\hat{d}^\dag_{k\mu\alpha}\hat{d}_{k\mu\alpha}
 \end{align}
with $\epsilon_{k\alpha}$ being the energy of an electron with wave vector $k$ in the $\alpha$ lead, and the $\hat{d}^\dag_{k\mu\alpha}$($\hat{d}_{k\mu\alpha}$) corresponding creation (annihilation) operator for an electron with the $\alpha$-reservoir state $|k\rangle$ of energy $\epsilon_{k\alpha}$.
 The dot-electrode coupling Hamiltonian is
 \begin{align}\label{hd}
     H_{SB}=\sum_{\mu}[f^\dag_{\mu}(t)\hat{a}_{\mu} +\hat{a}^\dag_{\mu}f_{\mu}(t)]
  \end{align}
  in the bath interaction picture.  Here, $f^\dag_{\mu}=e^{iH_{B}t}[\sum_{k\alpha}t^{*}_{\alpha k
\mu}\hat{d}^\dag_{k\mu\alpha}]e^{-iH_{B}t}$ is stochastic
interactional operator and satisfies the Gauss statistics with
$t_{\alpha k\mu}$ denoting the transfer coupling matrix element.
The influence of  electrodes on the dots is considered through the hybridization functions with a Lorentzian form,
$\Delta_{\alpha}(\w)\equiv\pi\sum_{k} t_{\alpha k\mu}t^\ast_{\alpha k\mu} \delta(\w-\ep_{k\alpha})=\Delta
W^{2}/[2(\w-\mu_{\alpha})^{2}+W^{2}]$, with $\Delta$ being the
effective impurity-lead coupling strength, $W$ being the band width,
and $\mu_{\alpha}$ being the chemical potentials of the
$\alpha$ lead.

Obviously, \Eq{ha} is a strongly correlated Hamiltonian with infinite degree of freedoms, which is hard to exactly solve by the the Schr\"{o}dinger equation directly. Fortunately, we can derive the accurate HEOM for the reduced density matrix (together with the auxiliary ones) from the basic path integral equations (influence functional theory) without take any approximations \cite{Jin08234703}.  The HEOM that governs the dynamics of the DQD takes the form of
\begin{align}\label{HEOM}
   \dot\rho^{(n)}_{j_1\cdots j_n} =& -\Big(i{\cal L} + \sum_{r=1}^n \gamma_{j_r}\Big)\rho^{(n)}_{j_1\cdots j_n}
     -i \sum_{j}\!     
     {\cal A}_{\bar j}\, \rho^{(n+1)}_{j_1\cdots j_nj}
\nl &
    -i \sum_{r=1}^{n}(-)^{n-r}\, {\cal C}_{j_r}\,
     \rho^{(n-1)}_{j_1\cdots j_{r-1}j_{r+1}\cdots j_n}
\end{align}
where the $n$th-order auxiliary density operator $\rho^{(n)}$ can be defined via auxiliary influence functional $\mathcal{F}^{(n)}_{\textbf{j}}$ as
\begin{align}\label{HEOM}
   \rho^{(n)}_{\textbf{j}}(t)\equiv \mathcal{U}^{(n)}_{\textbf{j}}(t,t_{0})\rho(t_{0})
\end{align}
with the reduced Liouville-space propagator,
\begin{align}\label{HEOM}
   \mathcal{U}^{(n)}_{\textbf{j}}(\psi,t;\psi_{0},t_{0}) \equiv \int^{\psi[t]}_{\psi_{0}[t_{0}]}\mathcal{D}\psi e^{i\mathcal{S}[\psi]} \mathcal{F}^{(n)}_{\textbf{j}}[\psi] e^{-i\mathcal{S}[\psi^{'}]}
\end{align}
$\mathcal{S}[\psi]$ is the classical action functional of the reduced system. The definition of the auxiliary influence functional $\mathcal{F}^{(n)}_{\textbf{j}}$ together with its equations is referred to in \cite{Jin08234703}.

We denote $\textbf{j}=\{j_1\cdots j_n\}$ and
$\textbf{j}_{r}=\{j_1\cdots j_{r-1}j_{r+1}\cdots j_n\}$, the action
of superoperators respectively is
\begin{align}\label{HEOM}
   {\cal A}_{\bar j}\, \rho^{(n+1)}_{\textbf{j}j} = a^{\bar{o}}_{\mu} \rho^{(n+1)}_{\textbf{j}j}
+(-)^{n+1} \rho^{(n+1)}_{\textbf{j}j}a^{\bar{o}}_{\mu}
\end{align}
\begin{align}\label{HEOM}
  {\cal C}_{j_r}\,\rho^{(n-1)}_{\textbf{j}_{r}} = \sum_{\nu} \{{\cal C}^{o}_{\alpha\mu\nu} a^{o}_{\nu}\rho^{(n-1)}_{\textbf{j}_{r}}
 &
    -(-)^{n-1}\, {\cal C}^{\bar{o}}_{\alpha\nu\mu}\,\rho^{(n-1)}_{\textbf{j}_{r}}a^{o}_{\nu}\}
\end{align}
In this formalism, $a_{\mu}^{o}$ ($a_{\mu}^{\bar{o}}$) corresponds the creation (annihilation) operator for an electron with the $\mu$ electron orbital. The reduced system density operator
$\rho^{(0)}(t) \equiv {\rm tr}_{B}[\rho_{\T}(t)]$ and auxiliary
density operators $\{\rho^{(n)}_{j_1\cdots j_n}(t); n=1,\cdots,L\}$
are the basic variables, here $L$ denotes the terminal or truncated
tier level.
The Liouvillian of dots, $\mathcal{L}\,\cdot \equiv \hbar^{-1}[H_{\s}, \cdot\,]$,
contains the  {\it e-e} interactions. The index $j \equiv (o\mu m)$ corresponds to the
transfer of an electron to/from ($o=+/-$) the impurity state
$|\sigma\rangle$, associated with the characteristic memory time
$\gamma_m^{-1}$. The correlation function ${\cal
C}^{o}_{\alpha\mu\nu}(t-\tau)=\langle
f^{o}_{\alpha\mu}(t)f^{\bar{o}}_{\alpha\nu}(\tau)\rangle_{B}$ follows immediately the time-reversal symmetry and detailed-balance relations.

We set the initial total system at equilibrium where
$\mu_{\alpha}=\mu^{eq}=0$. The system will leave
equilibrium after applying a voltage to the left (L)
and right (R) leads,  and there will be a current flowing into the
$\alpha$-lead $I_{\alpha}(t)$
\begin{align}\label{hd}
    I_{\alpha}(t)=i\sum_{\mu}\mathrm{tr}_{s}[{\rho^\dag_{\alpha \mu}(t)\hat a_{\mu} -\hat a^\dag_{\mu}\rho^-_{\alpha \mu}(t)}]
  \end{align}%
Here, $\rho^\dag_{\alpha \mu}=(\rho^-_{\alpha \mu})^\dag$ is the
first-tier auxiliary density operator
obtained by solving Eq.(5). Through the extended Meier-Tannor parametrization
method and multiple-frequency-dispersed hierarchy
construction, we can achieved the closed HEOM formalism \cite{Jin08234703}. As a result, the current from $L$ to $R$
lead can be denoted
$I(t)=I_{L}(t)=-I_{R}(t)$.

For evaluation of dynamical variables of the DQD system, we
focus on correlation function between two arbitrary dynamical
operators, $\widetilde{C}_{AB}(t) \equiv \la \hat A(t)\hat B(0)\ra =
{\rm tr}_{\T}[\hat A(t)\hat B(0) \rho^{\rm eq}_{\T}(T)]$.
Here, the Heisenberg operators and thermal equilibrium density operator $\rho^{\rm eq}_{\T}(T)$ are all defined in the total space. A linear response theory for quantum open systems \cite{Wei2011arxiv,Li2012dresponse} has been established, based on which $\widetilde{C}_{AB}(t)$ is retrieved exactly within the HEOM framework.
Let $C_{AB}(\w) \equiv \frac{1}{2}\int dt\, e^{i\w t}
\widetilde{C}_{AB}(t)$, which satisfies the detailed balance
relation of $C_{BA}(-\omega)=e^{-\hbar\omega/k_B T}C_{AB}(\omega)$.
The system spectral function is obtained as
$J_{AB}(\w)  \equiv \frac{1}{2\pi} \int dt\, e^{i \w t}
 \langle \{ \hat A(t), \hat B(0) \} \rangle
 = \frac{1}{\pi} \left(1 + e^{-\hbar\omega/k_B T}\right) C_{AB}(\w)$.
With $\hat A=\hat a_{\mu}$ and $\hat B=\hat a^{\dg}_{\mu}$, it
recovers the spectral function of the impurity state $\mu$,
\emph{i.e.}, $A_{\mu}(\omega) \equiv J_{\hat a_{\mu}\hat
a^{\dg}_{\mu}}(\omega) = -\frac{1}{\pi}{\rm Im}\,
G_{\mu\mu}(\omega)$. Here, $G_{\mu\mu}(\omega)$ is the retarded Green's function.

In our calculations, we treat the results as converging if the errors in numerical results of each element of the density matrix or the matrix of spectral function between the truncation $L=N$ and $L=N+1$ are less than $5\%$,  then sufficiently accurate current will be output. In the follow calculations we adopt $L=4$ .

 The main advantages of the HEOM approach applying to the DQD systems are as follows: 1) the HEOM theory is established based on the Feynman-Vernon path-integral formalism, in which all the system-bath correlations are taken into consideration; 2) the HEOM method is nonperturbative. In principle, the HEOM formalism is formally exact for noninteracting electron reservoirs. It also resolve nonperturbatively the combined effects of \emph{e-e} interactions; 3) the HEOM is a high-accuracy numerical approach. It has the ability to achieve the same level of accuracy as the latest high-level NRG method \cite{Li2012dresponse}.  Its main disadvantage lies in  the increasing computational cost as the system temperature decreases.

\section{\label{Rest}RESULTS AND DISCUSSION}
\subsection{\label{Pict}Physical picture of Pauli spin blockade}
\begin{figure}[!ht]
\centering
\includegraphics [width=3.5in]{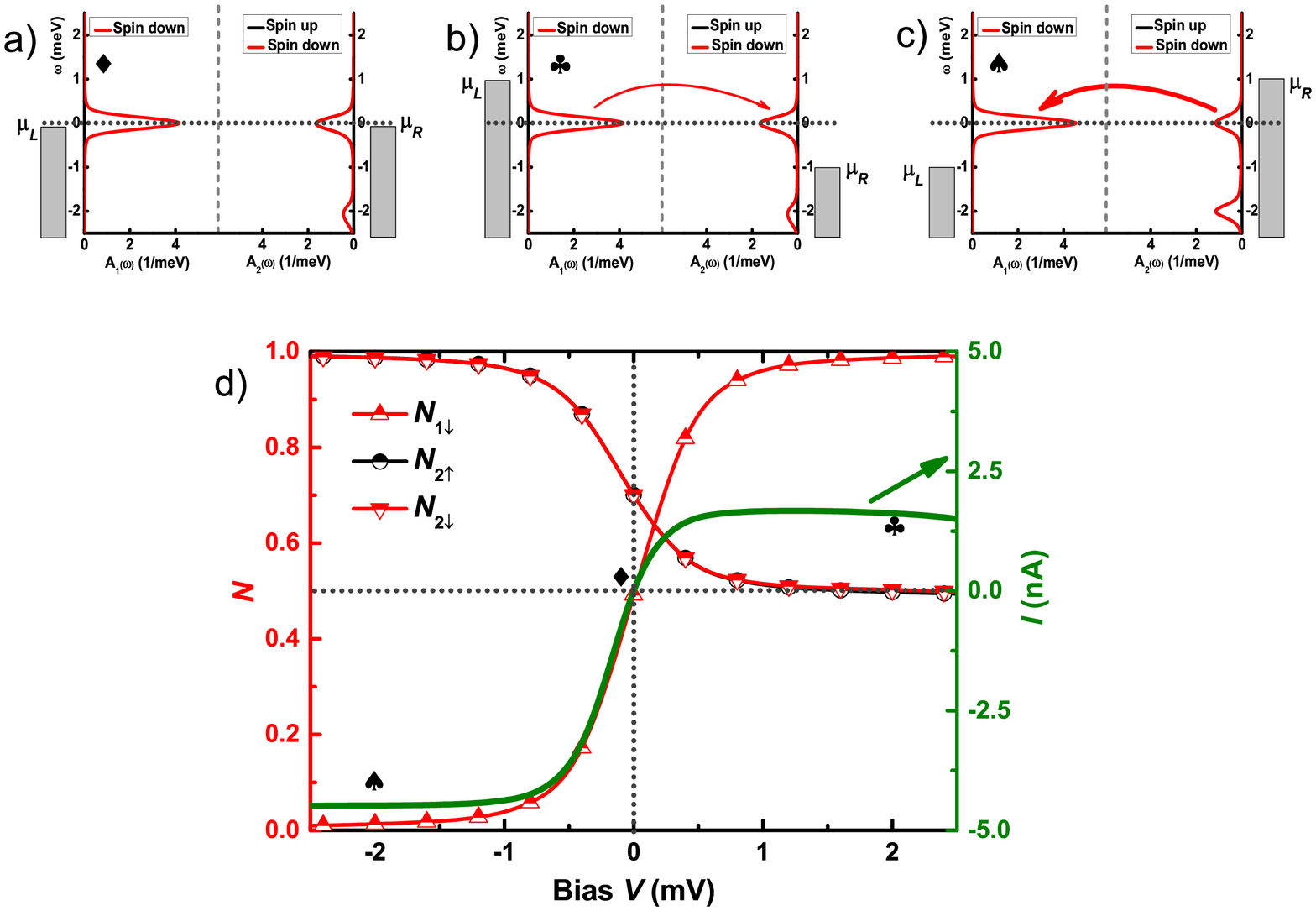}
\caption{(Color online). In the case of the spin non-degeneracy in dot 1, (a) equilibrium($V=0$) spectral function $A_{1\downarrow}(\omega)$, $A_{2\uparrow}(\omega)$ and $A_{2\downarrow}(\omega)$; (b) nonequilibrium spectral function $A_{1\downarrow}(\omega)$, $A_{2\uparrow}(\omega)$ and $A_{2\downarrow}(\omega)$ at positive bias($V=+2$ mV);(c) nonequilibrium spectral function $A_{1\downarrow}(\omega)$, $A_{2\uparrow}(\omega)$ and $A_{2\downarrow}(\omega)$ at negative bias ($V=-2$ mV); and (d) the dependence of the occupation numbers $N_{1\downarrow}$, $N_{2\uparrow}$, $N_{2\downarrow}$ as well as the total current $I$ on the bias voltage $V$. The parameters adopted are  $\varepsilon_{1\uparrow}=\varepsilon_{1\downarrow}=\varepsilon_{2\uparrow}
=\varepsilon_{2\downarrow}=-1.0$,  $U_{1}=U_{2}=2.0$,  $V_{g1}=-2.5$, $V_{g2}=1.0$,  $W_{L}=W_{R}=4.0$, $\Gamma_{L}=\Gamma_{R}=2.0$, $\Delta E_{B1}=1.5$, $T=0.1$ and $t=0.001$ (in unit of meV).}\label{fig1}
\end{figure}

The spin degeneracy in \Eq{ha} is not convenient for the detailed analysis of the PSB, thus we theoretically lift the spin degeneracy in dot 1 by means of a local  magnetic field $B_{1}$ applied onto it, with its direction paralleling to the down spins.  $B_{1}$ is chosen to be strong enough to make the energy level of up-spin electrons $\epsilon_{1\uparrow}$ much higher than $\epsilon_{1\downarrow}$. In experiments, a local-like inhomogeneous Zeeman field can be archived by a novel split micromagnet \cite{Brun2011dqbit}. By adjusting gate voltage $V_{1}$, we then set $\epsilon_{1\downarrow}$ at the equilibrium Fermi level $E_{F}$ ($E_{F}=\mu_{L}=\mu_{R}$ at zero-bias).  Under such condition, $E_{F}$ coincides with the center of the peak of the transition for down-spin electrons in dot 1 to jump from the zero occupied level to single occupied one, as the spectral function $A_{1\downarrow} (\omega)$ shown in \Fig{fig1}(a). This kind of single-occupied transition peak of $A_{1\uparrow} (\omega)$ is pushed far higher than  $E_{F}$ by the large $B_{1}$,  which makes the up-spin current
negligibly small, thus $A_{1\uparrow} (\omega)$ is not shown in figures [see \Fig{fig1}(a)-(c)] but its little contribution to the total current is still counted [see \Fig{fig1}(d)].

Without the coupling $t$ between dot 1 and 2, the local  $B_{1}$ should take no direct effects on the spins in dot 2. By adjusting gate voltage $V_{2}$, we set the double-occupied transition (from the single occupied level to double occupied one) peaks of $A_{2\uparrow}(\omega)$ and $A_{2\downarrow}(\omega)$ to coincide with  $E_{F}$ at $t\sim0$, as shown in \Fig{fig1}(a). Proceeding from this set-up, we gradually adjust the inter-dot coupling strength and other parameters to elaborate the physical picture of the PSB in details.

We start our study on the PSB from the following parameters: the potential energy $\varepsilon_{1\uparrow}=\varepsilon_{1\downarrow}=\varepsilon_{2\uparrow}
=\varepsilon_{2\downarrow}=-1.0$ meV; on-site {\it e-e} interaction strength $U_{1}=U_{2}=2.0$ meV; gate voltage $V_{g1}=-2.5$ meV and $V_{g2}=1.0$ meV. The left and right electrodes are chosen to be symmetric, with their DOS being Lorentzian-type; bandwidth being $W_{L}=W_{R}=4.0$ meV and the electrode-dot coupling being $\Gamma_{L}=\Gamma_{R}=2.0$ meV. The Zeeman splitting energy caused by the local magnetic field $B_{1}$ is $\Delta E_{B1}=1.5$ meV; and the temperature $T=0.1$ meV.  The dot-dot coupling in \Fig{fig1} is near zero, $t=0.001$ meV.

The spectral functions $A(\omega)$ shown in \Fig{fig1}(a) is calculated by HEOM at the equilibrium state (bias voltage $V=0$), where only $A_{1\downarrow}(\omega)$ is shown in dot 1.  \Fig{fig1}(b) and (c) show nonequilibrium spectral functions $A_{1\downarrow}(\omega)$, $A_{2\uparrow}(\omega)$ and $A_{2\downarrow}(\omega)$ at positive ($V=+2$ mV) and negative ($V=-2$ mV) bias, respectively.  \Fig{fig1}(d) depicts the dependence of the occupation numbers $N_{1\downarrow}$, $N_{2\uparrow}$, $N_{2\downarrow}$ as well as the total current $I$ on the bias voltage $V$ ($N-V/I-V$ curves), where $N$ is obtained from the diagonal elements of the reduced density matrix,  which is also equal to the weighted integral of corresponding $A(\omega)$ under the Fermi level at the steady state, $\rho=\frac{1}{2\pi}\int A(\omega)f(\omega)d\omega$.

As shown in \Fig{fig1}, without the bias being applied ($V=0$), $N_{1\downarrow}\approx0.5$, $N_{2\uparrow}=N_{2\downarrow}\approx0.7$.  The fractional charges  results from the following two reasons: 1) the value of the Fermi-Dirac function $f(\omega)$ in the vicinity of $E_F$ continuously changes from 0 to 1 rather than a integer; and 2) the spectral function $A(\omega)$ near $E_F$ shows a finite-width peak structure broadened by the dot-electrode interaction, thus the weighted integral value below the Fermi surface is less than 1. The total electron number in the two-dot system is $Q_{T}=N_{1\uparrow}+N_{1\downarrow}+N_{2\uparrow}+N_{2\downarrow}\approx2.0$ $(N_{1\uparrow}\sim0)$, which remains almost constant even in the non-equilibrium (steady-state) transport process if the external (electric and/or magnetic) field is not too large.

Then, the symmetrical positive ($\mu_{L}\rightarrow \mu_{L}+eV/2;\mu_{R}\rightarrow \mu_{R}-eV/2$) and negative ($\mu_{L}\rightarrow \mu_{L}-eV/2;\mu_{R}\rightarrow \mu_{R}+eV/2$) bias are applied to the configuration shown in  \Fig{fig1}(a). From the changes of $A(\omega)$, $N$ and $I$ with $V$ shown in \Fig{fig1}(b)-(d), we can see that the spectral function and occupation number of up-spin and down-spin electrons keep degenerate in dot 2,  and neither positive nor negative bias can lift the degeneracy. The reason lies in the very small coupling between two dots, $t\sim 0$.

The  $I-V$ curve [see \Fig{fig1}(d)] shows distinct asymmetric behavior that the steady value of the current  at $V<0$  is much larger than that at $V>0$, actually the former is almost twice of the latter. However, such asymmetry is not the same as the rectification since the positive steady current is not small enough (comparing to the negative one ) to define a blockage effect.  That result can be explained as follows:

In principle, the dot-to-dot electron transfer can induce an antiferromagnetic exchange between them with the strength $J_{AF}\sim 4t^2/U$. Enough strong $J_{AF}$ will lock the ground state of the isolate double dot system into a spin singlet state $S(1,1)$ whose energy is lower than the triplet state $T(1,1)$ in the order of $J_{AF}$.  In \Fig{fig1}, $t\sim 0$ means $J_{AF}\sim 0$, thus states of $S(1,1)$ and  $T(1,1)$ are nearly degenerate at $V=0$.

When positive bias applied, $\mu_{R}<0$, as shown in \Fig{fig1}(d), $N_{2\uparrow}$ and $N_{2\downarrow}$ decreases gradually with $\mu_{R}$ decreasing . The relation of $N_{2\uparrow}\approx N_{2\downarrow}$ at $V=0$ will keep unchanged at $V>0$, and the number will reach a steady value after $V>1$ mV, $N_{2\uparrow}\approx N_{2\downarrow}\approx 0.5$. At the same time,  $N_{1\downarrow}$ increases gradually with $\mu_{L}(\mu_{L}>0)$ increasing and tends to $1.0$ after $V>1$ mV. The nonequilibrium spectral function at $V=+2$ mV in \Fig{fig1}(b) further confirms the above process. It means that the states $S(1,1)$ and $T(1,1)$ remain degenerate at $V>0$, and each of them has $50\%$ probability after $V>1$ mV. It is believed that electron spin is conserved through direct hopping process,thus the initial state $T(1,1)$ can only transfer to  $T(0,2)$, if one electron is driven from dot 1 to dot 2 by the positive bias. However, the state $T(0,2)$ is not permitted to exist due to the Pauli's exclusion principle (the exception from the excess freedom such as the orbital or valley not considered in the present work). Therefore,  only half of the total initial states [$S(1,1)$] can contribute to the transport current at $V>0$ via the transition $S(1,1)\rightarrow S(0,2)$.

As shown in \Fig{fig1}(d), the negative bias makes $\mu_{R}>0$, then $N_{2\uparrow}$ and $N_{2\downarrow}$ increases gradually with $\mu_{R}$ increasing but remains  $N_{2\uparrow}\approx N_{2\downarrow}$ which is close to $1.0$ after $V<-1$ mV. Meanwhile, $N_{1\downarrow}$ decreases gradually with $\mu_{L}(\mu_{L}<0)$ decreasing and approaches to $0$ after $V<-1$ mV.  The nonequilibrium spectral function at $V=-2$ mV in \Fig{fig1}(c) further confirms the above process. It suggests that the negative bias will stabilize the $S(0,2)$ state which  $100\%$ contributes to the current, since the Pauli's exclusion principle takes no effect on the transition from $S(0,2)$ to $S(1,1)$. That argument is verified by the $I-V$ curve in \Fig{fig1}(d), which shows that the steady current at $V<0$ is almost twice larger than that at $V>0$.

\begin{figure}[!ht]
\centering
\includegraphics [width=3.5in]{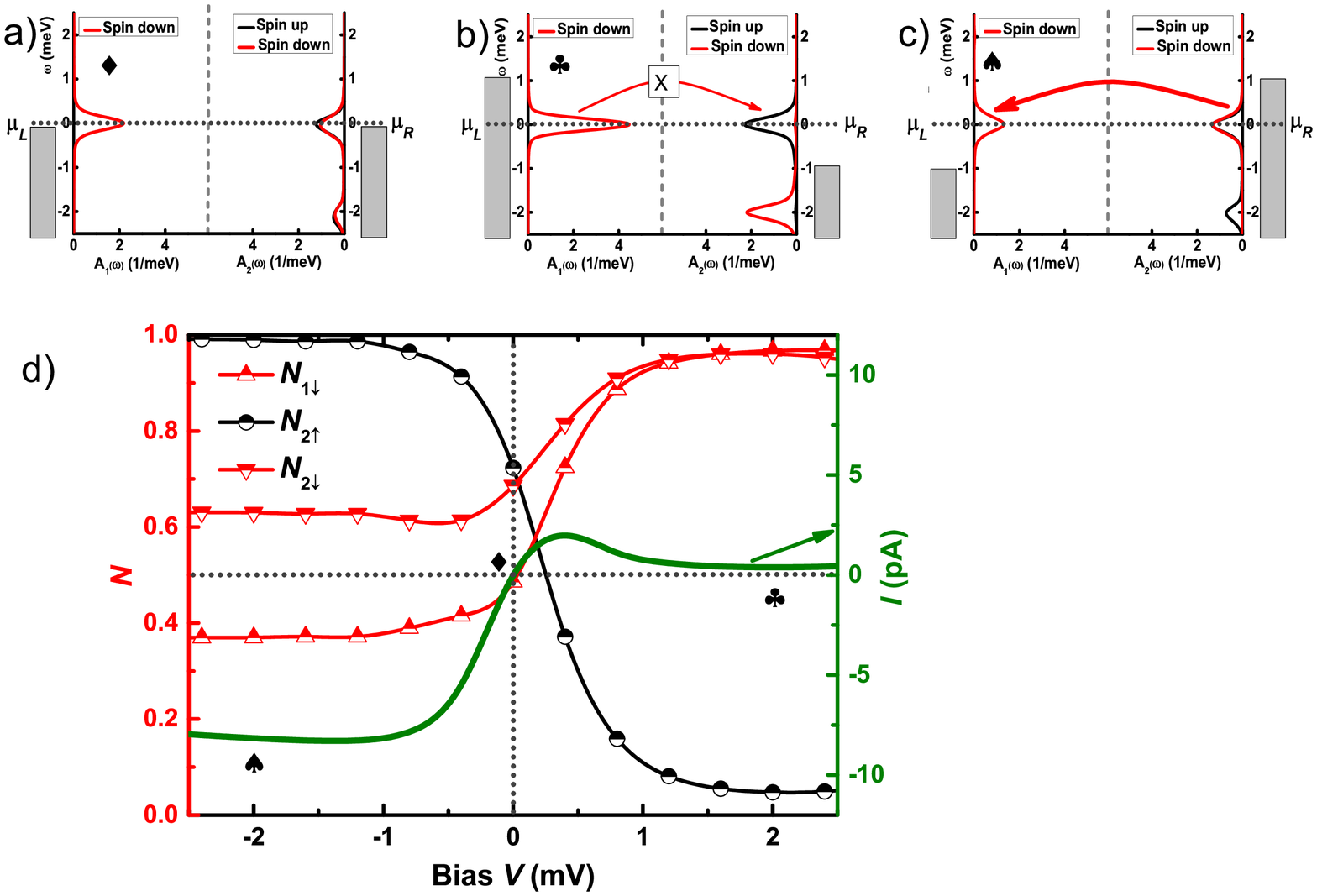}
\caption{(Color online). In the case of the spin non-degeneracy in dot 1, (a) equilibrium($V=0$) spectral function $A_{1\downarrow}(\omega)$, $A_{2\uparrow}(\omega)$ and $A_{2\downarrow}(\omega)$; (b) nonequilibrium spectral function $A_{1\downarrow}(\omega)$, $A_{2\uparrow}(\omega)$ and $A_{2\downarrow}(\omega)$ at positive bias($V=+2$ mV);(c) nonequilibrium spectral function $A_{1\downarrow}(\omega)$, $A_{2\uparrow}(\omega)$ and $A_{2\downarrow}(\omega)$ at negative bias ($V=-2$ mV); and (d) the dependence of the occupation numbers $N_{1\downarrow}$, $N_{2\uparrow}$, $N_{2\downarrow}$ as well as the total current $I$ on the bias voltage $V$. The parameters adopted are the same as those in \Fig{fig1}, except $t=0.08$  meV here.}\label{fig2}
\end{figure}

We then increase the inter-dot coupling strength $t$ from $0.001$ to $0.08$ meV and keep other parameters unchanged for the purpose of comparison. The calculated $A(\omega, V=0)$,  $A(\omega, V=+2$ meV$)$, $A(\omega, V=-2$ meV$)$ and the $N-V/I-V$ curve are depicted in \Fig{fig2}(a)-(d) respectively. From \Fig{fig2}(a) for the case of $V=0$, we can see that the degeneracy of $S (1,1)$ and $T(1,1)$ has been lifted by the $t$-induced antiferromagnetic interaction $J_{AF} (\sim 4t^{2}/U)$. Since the energy of $S(1,1)$ is lower than that of $T(1,1)$, $A_{2\uparrow}(\omega)$ moves downward and $N_{2\uparrow}$  increases from $\sim 0.7$ at $t\sim 0$ to $\sim 0.72$. Meanwhile, $N_{2\downarrow}$ decreases from $\sim 0.7$  to $\sim 0.68$, as shown in \Fig{fig2}(d).

The PSB will take place when the positive bias is applied to the set-up shown in \Fig{fig2}(a). At $V>0$, $N_{2\uparrow}$ and $N_{2\downarrow}$  changes to different directions [see \Fig{fig2}(d)] instead of synchronous varying at $t\sim 0$ [cf. \Fig{fig1}(d)]. As shown in \Fig{fig2}(d),  $N_{2\uparrow}$  decreases rapidly as $V$ positively increasing and tends to $0$ after $V>1.0$ mV; whereas $N_{2\downarrow}$  increases and approaches $1.0$ after $V>1.0V$ mV. As for $N_{1\downarrow}$, it  gradually increases from $0.5$ to $1.0$ with increasing $V$, much like the change of  $N_{2\uparrow}$.  As a result of above changes, only state $T(1,1)$ is retained under positive bias and  $S(1,1)$ will no longer exist after $V>1.0$ mV. As already explained, $T(1,1)$ can not transfer to $T(0,2)$ to create any current due to the  Pauli's exclusion principle, thus PSB occurs naturally. The forbidden electron transition from dot 1 to 2 outputs near zero current, as the $I-V$ curve shows in \Fig{fig2}(d).  That is exactly the PSB effect observed in experiments at a moderate coupling strength $t$, and the finite current in the interval of $V\in[0,1.0$ mV$]$ is the so-called leakage current corresponding to the process of  $S(1,1)$ being depleted gradually. After $V>1.0$ mV, the current enters its total-blocked zone with a near zero value.

When negative bias applied as shown in \Fig{fig2}(d), with the increase of $\mu_{R}$, $N_{2\uparrow}$ increases gradually and becomes saturated at $N_{2\uparrow}\sim 1.0$ after the bias $V<-1.0$ mV, while $N_{2\downarrow}$ decreases and keeps at about $ 0.5+\delta$ ($\delta\approx0.13$) after  $V<-1.0$ mV. At the same time, $N_{1\downarrow}$ decreases with $\mu_{L}$ decreasing and maintains $N_{1\downarrow}\approx 0.5-\delta$ after $V<-1.0$ mV. Although  $N_{1\downarrow}>0$ in this case, it won't cause any PSB effect of the down-spin electrons, for the reason that the $\mu_{L}$ below the Fermi surface can provide enough space to accept the electrons transferring from dot 2 to 1, as the nonequilibrium spectral function shown in \Fig{fig2}(c). As a consequence, the considerable current is output in the $I-V$ curve. It should be noted that the unit of current in \Fig{fig2}(d) is pA instead of nA in \Fig{fig1}(d).

\begin{figure}[!ht]
\centering
\includegraphics [width=2.3in]{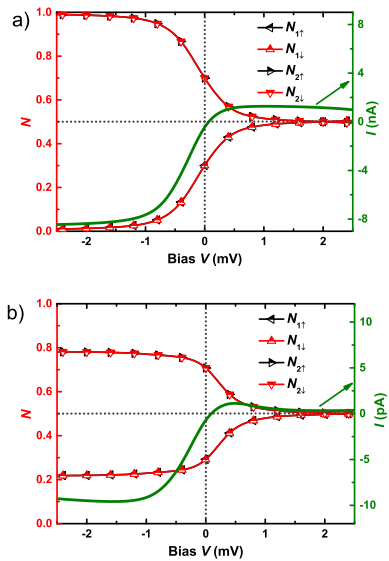}
\caption{(Color online). Without the local magnetic field applied ($B_1=0$), the dependence of the occupation numbers$N_{1\uparrow}$, $N_{1\downarrow}$, $N_{2\uparrow}$, and $N_{2\downarrow}$ as well as the total current $I$ on the bias voltage $V$ at $t=0.001$ meV (a) and $0.08$ meV (b). The other parameters in (a) and (b) are the same as those in \Fig{fig1}  and \Fig{fig2} accordingly.}\label{fig3}
\end{figure}

Summarizing \Fig{fig1} and \ref{fig2}, our theory appropriately describes the physical mechanism and picture of the PSB effect in DQD systems, by means of a strong local magnetic field applied onto dot 1 to lift its spin degeneracy.  We are now on the position to elucidate the PSB effect under more general conditions. In \Fig{fig3}, we depict the $N-V/I-V$ curves at $t=0.001$ meV [\Fig{fig3}(a)] and $0.08$ meV [\Fig{fig3}(b)], where the other parameters are the same as those in \Fig{fig1} and \Fig{fig2} accordingly, except that the local magnetic field is absent in both cases, i.e. $B_1=0$. By comparing the $I-V$ curve in \Fig{fig3}(a) to that in \Fig{fig1}(d), we can see that the additional up-spin channel barely changes the current at $V>0$ but will increase that at $V<0$ if the dot-dot coupling is very weak. The reason for the former lies in the equally dividing of $N_{1\downarrow}$ in \Fig{fig1}(d) by $N_{1\uparrow}$ and $N_{1\downarrow}$ in \Fig{fig3}(a) and the current changing little. The reason for the latter is that the up- and down-spin channels contribute to the current independently in the limit of $t\approx 0$, thus the current will be enhanced by additional channels. If distinct PSB effect occurs, the situation will become much different. As indicated by the comparison of the $I-V$ curve in \Fig{fig3}(b) to that in \Fig{fig2}(d),  the additional up-spin channel hardly changes the current either in the PSB region ($V>0$) or in the conductive region ($V<0$). By analyzing the corresponding $N-V$ curves and nonequilibrium spectral functions (the figures not shown), we find in the PSB region, the probability of $T(\downarrow, \downarrow)$ in \Fig{fig2}(d) is equally divided into $T(\uparrow,\uparrow)$ and $T(\downarrow,\downarrow)$ in \Fig{fig3}(b), and the current keeps its value.  In the conductive region,  the single-spin transport channel $S(0,\uparrow \downarrow)\rightarrow S(\downarrow,\uparrow)$ is equally divided by the degenerate double-spin ones,  $S(0,\uparrow \downarrow)\rightarrow \frac{1}{2}S(\downarrow,\uparrow)+\frac{1}{2}S(\uparrow,\downarrow)$, and the total current remain unchanged.

\subsection{\label{Gmodu}Gate voltage modulation on PSB}

\begin{figure}[!ht]
\centering
\includegraphics [width=2.5in]{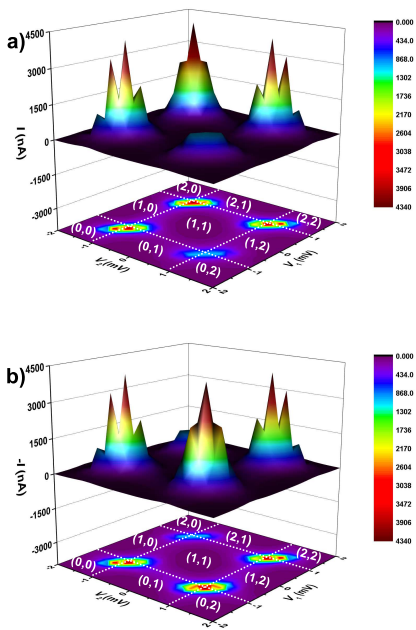}
\caption{(Color online). 3D colormap surface image with the 2D bottom contour projection of the current (absolute value) as a function of the gate voltage $V_{1}$ and $V_{2}$. The bias voltages used are (a) $V=0.4$ mV, and (b) $V=-0.4$ mV. In the 2D projection image, the dotted lines schematically mark off the boundary of the stability diagrams. }\label{fig4}
\end{figure}

In experiments on  PSB, the gate voltages $V_{1}$ and $V_{2}$ are two important parameters which respectively manipulate the on-site energy of dot 1 and dot 2. We thus theoretically investigate the variation of the PSB with them and summarize the results in \Fig{fig4}, where the dot-dot coupling strength is chosen as $t=0.05$ meV, a small value but large enough to induce the PSB effect, with the purpose of making the boundaries shown in \Fig{fig4} clear and distinguishable.  \Fig{fig4} (a) depicts the positive current at $V=0.4$ mV changing with respect to the parameters in the $V_{1}-V_{2}$ plane, in the form of the 3D colormap surface image together with the 2D bottom contour projection. The other parameters are the same as those in \Fig{fig3}, except both $V_{1}$ and $V_{2}$ are variables now within the range of $[-U,U]$($U=2$ meV). In the 2D projection image, we schematically mark off the boundary of the stability diagrams by dotted lines.  Actually, the quadrangles shown in the figure should changes to hexagon at finite $t$, however, the boundary line is hard to accurately determine in theory. The schematic stability diagrams shown in \Fig{fig4} is just for reference purposes.

\Fig{fig4} clearly shows how ($V_{1},V_{2}$) modulate ($N_1,N_2$)  and the PSB effect. The DQD system remains the stability charge-occupied state (1,1) within $-U/2\leq V_{1}\leq U/2$ and $-U/2\leq V_{2}\leq U/2$, and no current occurs in the center of this area due to the Coulomb blockade.  However, finite current may be output at the four top corners of (1,1) state, resulting from the charge transferring in and out, which can be listed as  $(0,0)\rightleftharpoons (1,1), (2,0)\rightleftharpoons (1,1),(1,1)\rightleftharpoons (2,2)$, and $(1,1)\rightleftharpoons (0,2)$ (clockwise from lower left). By referring the figure, one can see that the current at the corners of $(0,0)\rightleftharpoons (1,1)$ and $(1,1)\rightleftharpoons (2,2)$ is symmetric about the bias voltage with no PSB effect occurring.  In addition, the current at above two corners also shows symmetric behaviors along the diagonal line ($V_{1}=V_{2}$), which comes from the electron-hole symmetry satisfied by our Hamiltonian, \Eq{ha}. As shown in \Fig{fig4}, current at the corners of $(2,0)\rightleftharpoons (1,1)$ and $(1,1)\rightleftharpoons (0,2)$ exhibits rectifying characters about positive and negative bias. The former case has been elaborated in \Fig{fig1} to \ref{fig3}, and the small current ($|I|<1$ pA) shown in \Fig{fig4}(b) corresponds to the leakage current at low bias shown in \Fig{fig3}. The latter case of $(1,1)\rightleftharpoons (0,2)$ is very similar except that the PSB effect takes place at the positive bias.

\begin{figure}[!ht]
\centering
\includegraphics [width=3.8in]{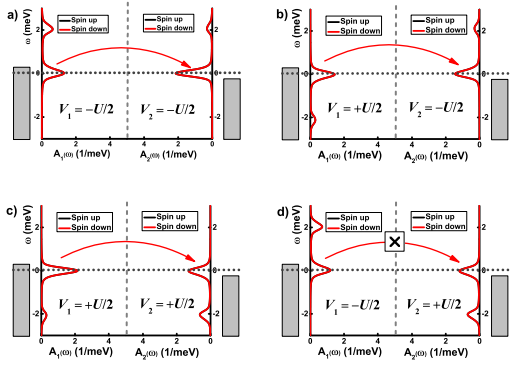}
\caption{(Color online). Non-equilibrium spectral functions at four top corners of (1,1) state in \Fig{fig4}. (a), (b), (c) and (d) respectively represent the point $(-U/2,-U/2), (U/2,-U/2)$, $(U/2,U/2)$ and $(-U/2,U/2)$ in the $(V_{1},V_{2})$ parameter plane.}\label{fig5}
\end{figure}

By referring \Fig{fig4}, one can see that large current more than 4 pA can occur at the top corners of $(1,1)$ state, namely the center points of transition between $(1,1)$ and other stability states. In \Fig{fig4}(a), those points correspond to $(-U/2,-U/2), (U/2,-U/2)$ and $(U/2,U/2)$ in the $(V_{1},V_{2})$ parameter plane, versus $(-U/2,-U/2), (-U/2,U/2)$ and $(U/2,U/2)$ in \Fig{fig4}(b). What special about those points is that one quantum transition peak in dot 1 will resonate with another one in dot 2 at the Fermi surface, as the nonequilibrium spectral functions shown in \Fig{fig5}, where \Fig{fig5}(a) to (d) corresponds to the point $(-U/2,-U/2)$ to $(-U/2,U/2)$ in the clockwise order. Taking \Fig{fig5}(a) as an example, we can see that the resonance on point $(-U/2,-U/2)$ takes place between the transition of $(0\rightarrow 1)$ in dot 1 and that of $(0\rightarrow 1)$ in dot 2 at the Fermi surface,  which induces a very large current, $I\sim4.2$ pA. If  ($V_{1},V_{2}$) deviates those points parallel to  the $V_{1}=V_{2}$ diagonal line, the resonance between the transition peaks will still exist, but no longer coincide with the Fermi surface. As a consequence, the current will gradually decrease into the Coulomb blockade region, after some peak-like structures, as shown in the 3D colormap surface image in \Fig{fig4}. If  ($V_{1},V_{2}$) deviates the four top corners of (1,1) state along any direction of $V_{1}\neq V_{2}$, none of the resonance will survive, and then the current will decays to very small value ($I<0.5$ pA) quickly.

Although both peaks of the transition of $(0\rightarrow 1)$ in dot 1 and that of $(1\rightarrow 2)$ in dot 2 are coincide with the Fermi surface on the point $(-U/2,U/2)$, the PSB prohibits the resonance between them as shown in \Fig{fig5}(d), and only very small leakage current ($I<1$ pA) occurs in the vicinity of $(-U/2,U/2)$ [see \Fig{fig4}(a)]. When the direction of the bias is reversed from $V>0$ to $V<0$,  $(-U/2,U/2)$ will change from a PSB point to a resonance one, and then very large current occurs at this point. Accordingly, the PSB point moves to $(U/2,-U/2)$, as shown in \Fig{fig4}(b).

As indicated in \Fig{fig4}, our theoretical results of the gate voltage modulation on PSB is qualitatively consistent with the experimental measurements. The shape of conduct regions is also similar to the triangles observed in experiments, but not exactly the same. The difference may come from the reason that the Anderson multi-impurity model can not describe all the details in experiments. For instance, when the gate voltage $V_{1}$ on dot 1 changes, it has been confirmed by experiments that the effects on dot 2 are induced not only through the direct coupling $t$ but also through a capacitive coupling. We believe the former has been well described in our theory, but the latter has not yet.

\subsection{\label{Exag} Lift of PSB by exchange interaction}

Now we extend the Anderson two-impurity model to adding the term of dot-dot exchange interaction,  which describes the coupling between the local spins of two dots, with the Hamiltonian as follows:
\begin{equation}\label{EXH}
H_{E}=J\hat{S_{1}}\cdot\hat{S_{2}}
\end{equation}
where $\hat{S_{i}}$ is the spin operator of dot $i(i=1,2)$. $J$ is coupling strength between spins in different dots, which could be positive(antiferromagnetic) or negative(ferromagnetic).

Fundamentally, $J$ originates from the exchange term of the {\em e-e} interaction (potential Energy) between two dots, thus plays an equal important role as the kinetic energy $t$.  The manipulation of $J$ is a fascinating issue closely relevant to quantum information \cite{Hanson2007siqd}. Despite the practical difficulties, experiments may achieve this goal indirectly, for example, N. J. Craig {\em et al.} have demonstrated the control of the strength and the sign of $J$ between  two dots coupled through an open conducting region \cite{Craig2004RKKY}. In the present work, we investigate the effect of $J$ on PSB. When $t\neq 0$,  the total exchange interaction $J_{T}$ principally equals to the sum of $J$ and the antiferromagnetic one $J_{AF}$ induced by $t$, that is $J_{T}\approx J+4t^2/U$. In order to highlight the role of $J$, we thus choose a relatively small $t$ ($t=0.05$ meV)  at fixed $U=2.0$ meV to produce a very small $J_{AF}$ ($J_{AF}\sim 0.005$ meV).

\begin{figure}[!hbt]
\includegraphics [width=3.5in]{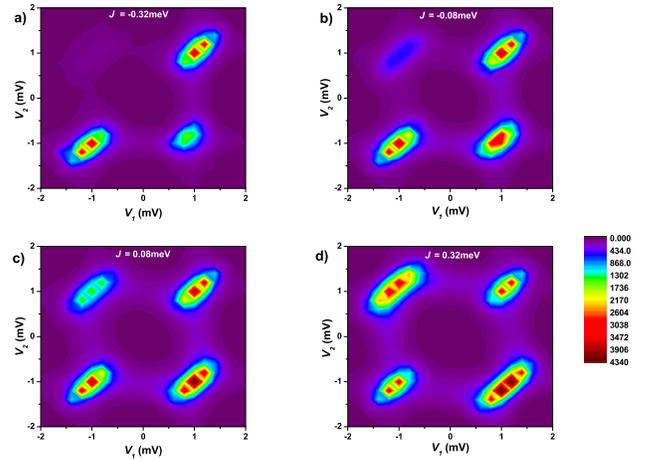}
\caption{(Color online). 2D  contour plot of the current (bias V=$0.4$ mV) as a function of the gate voltage $V_{1}$ and $V_{2}$ at various $J$. (a), (b), (c) and (d) correspond to $J=-0.32$, $-0.08$, $0.08$ and $0.32$ meV, respectively. }\label{fig6}
\end{figure}

\Fig{fig6} shows our results of gate voltage modulation (bias $V=0.4$ mV) on PSB at various $J$ with different signs and values, where  \Fig{fig6}(a), (b), (c) and (d) correspond to $J=-0.32$, $-0.08$, $0.08$ and $0.32$ meV, respectively.  By referring \Fig{fig6}, one can see that $J$ substantially affects the current near the points $(V_1, V_2)=(-U/2,U/2)$ and $(U/2,-U/2)$ where the PSB effects take place,  while it slightly dose to the current near $(-U/2,-U/2)$ and $(U/2,U/2)$ where no PSB effects. It indicates that the exchange interaction can directly change the characters of the PSB.

Let us focus on the change of the leakage current of the PSB around $(-U/2,U/2)$. It corresponds to a very small value at $J=-0.32$ meV [\Fig{fig6}(a)], which indicates that the ferromagnetic exchange interaction tends to enhance the PSB effect. Generally speaking, the leakage current will increase with the increasing of the algebra value of $J$, as shown in \Fig{fig6}. For example, the leak current increases from near zero to $\sim 0.5$ nA [\Fig{fig6}(b)] as $J$ increasing from $-0.32$ to $-0.08$ meV. Changing the sign of $J$ still maintains this kind of tendency, and the antiferromagnetic exchange interaction seems to suppress the PSB.  When  $J$ continually increases to $0.08$ meV [\Fig{fig6}(c)], the leakage current is clearly visible and in the range of $1\sim 2$ nA. If $J$ positively increases to a enough large value, e.g. $J=0.32$ meV [\Fig{fig6}(d)], the leakage current will increase distinctly, even to the same order of magnitude ($\sim 4$ nA) as the conductive current. In this case,  we argue that the PSB has been lifted by strong antiferromagnetic exchange interaction.

\begin{figure}[!hbt]
\includegraphics [width=3.5in]{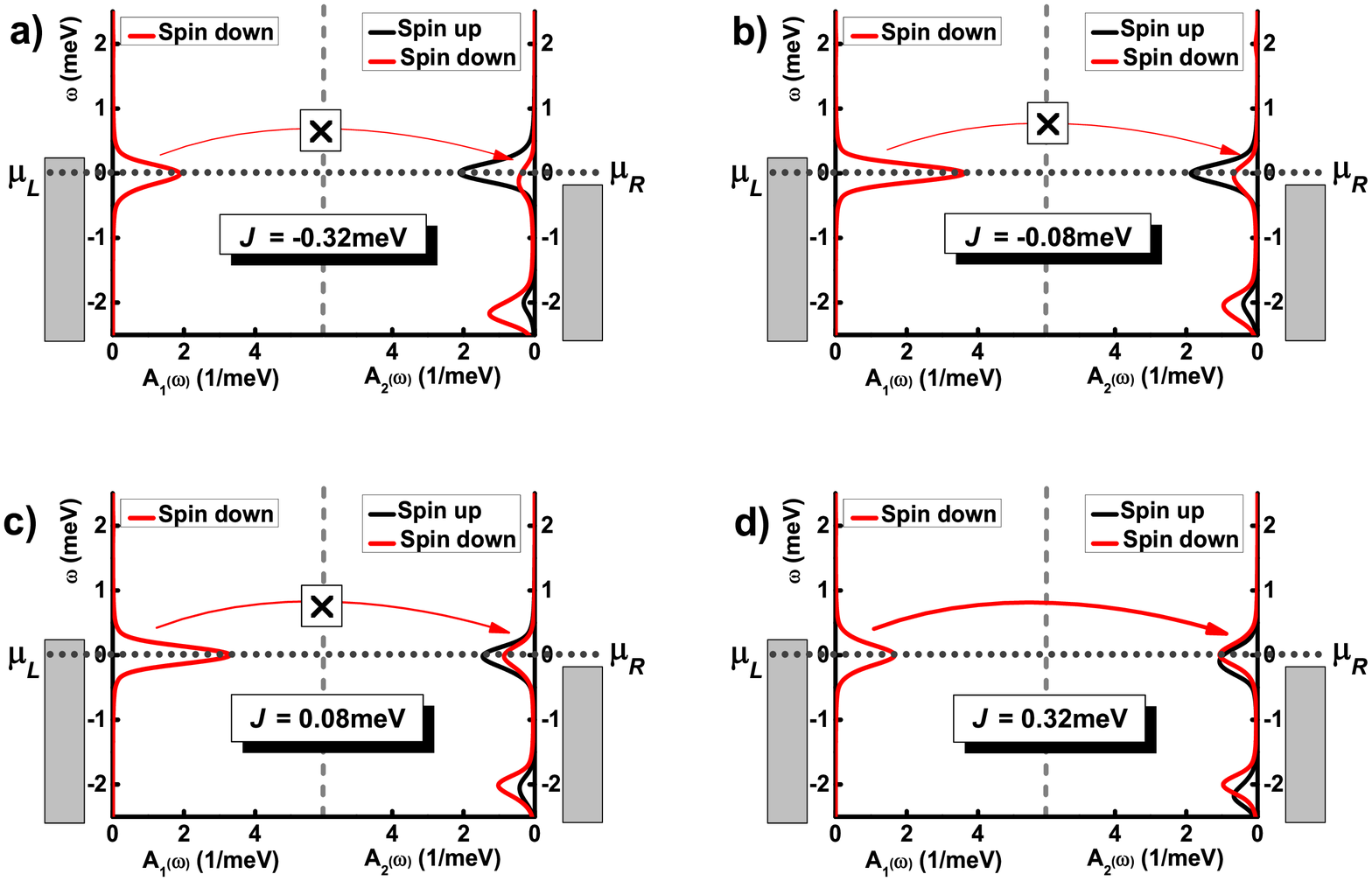}
\caption{(Color online). In the case of the spin non-degeneracy in dot 1, nonequilibrium (bias V=$0.4$ mV) spectral function $A_{1\downarrow}(\omega)$, $A_{2\uparrow}(\omega)$ and $A_{2\downarrow}(\omega)$ at $(V_1, V_2)=(-U/2, U/2)$. (a), (b), (c) and (d) correspond to $J=-0.32$, $-0.08$, $0.08$ and $0.32$ meV, respectively.}\label{fig7}
\end{figure}

The lift of the PSB is a significant feature of \Fig{fig6}, which deserves careful study to reveal its mechanism. We thus apply the local magnetic field $B_{1}$ again onto dot 1 to lift is spin degeneracy and investigate the transport of down-spin electrons in the PSB region. \Fig{fig7} shows the nonequilibrium spectral functions at the point of $(V_1, V_2)=(-U/2, U/2)$, with the Zeeman energy caused by $B_{1}$ being $\Delta E_{B1}=1.5$ meV and other parameters being the same as those in \Fig{fig6}. \Fig{fig7}(a)-(d) corresponds to \Fig{fig6}(a)-(d), respectively.

At $J=-0.32$ meV, the dot-dot exchange interaction is ferromagnetic, which makes the energy level of $T(1,1)$ lower than $S(1,1)$. As a consequence, the single occupied transition peak of down-spin in dot 2 is lower than that of up-spin, since that peak in dot 1 has been locked as down-spin one at the Fermi surface. It means that the down-spin electron in dot 2 has almost fully occupied the single level under the Fermi surface, which will prevent the hopping of electrons with the same spin and thus enhance the PSB effects. Continuously increasing the  algebra value of $J$ to $-0.08$ meV will not essentially change above process, and only slightly increase the weight of the down-spin holes in the  double occupied transition peak at the Fermi surface, which makes the leakage current increase to  a small nonzero value, as shown in \Fig{fig6}(b).

 When $J>0$, the spin exchange interaction is anti-ferromagnetic, which makes the energy level of $S(1,1)$ is lower than that of $T(1,1)$. This case should help electrons in dot 1 transfer to dot 2 by means of the transition $S(1,1)\rightarrow S(0,2)$, so the leak current will increase, which has been confirmed by the change of the spectral functions shown in \Fig{fig7}(c) and \Fig{fig7}(d). We can see that after the sign change of $J$, the weight of the down-spin holes in the double occupied transition peak increases further, offering more space for down-spin electrons to transfer from dot 1, and consequently increases the leak current value. Since $J$ is small in \Fig{fig7}(c),  the leak current can only increase to a relatively small value, as shown in \Fig{fig6}(c). In \Fig{fig7}(d), large enough $J$ ($J=0.32$ meV) makes the weight of down-spin hole increase remarkably, thus induces the large leak current shown in \Fig{fig6}(d) which lifts the PSB effect. By referring \Fig{fig6} and \ref{fig7}, we can see that the lift of PSB by $J$ is a continuous process instead of a sudden change.

 \begin{figure}[!hbt]
\includegraphics [width=2.5in]{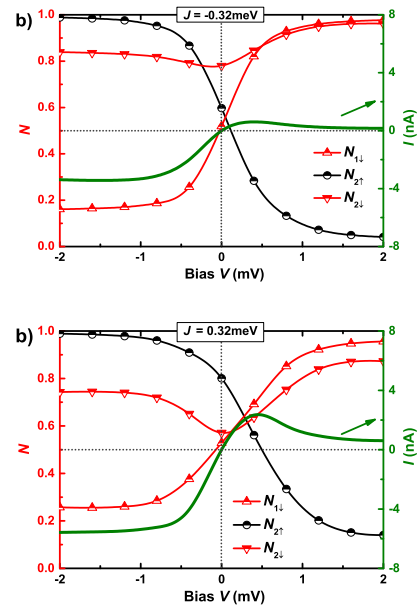}
\caption{(Color online). In the case of the spin non-degeneracy in dot 1, the dependence of the occupation numbers $N_{1\downarrow}$, $N_{2\uparrow}$, and $N_{2\downarrow}$ as well as the total current $I$ on the bias voltage $V$ at $J=-0.32$ meV (a) and $0.32$ meV (b).}\label{fig8}
\end{figure}

To further verify the mechanism of the lift of PSB by the exchange interaction, we calculate the $N-V/I-V$ curves at different $J$ and summarize the results in \Fig{fig8}, where \Fig{fig8}(a) corresponds to $J=-0.32$ meV and \Fig{fig8}(b) to $J=0.32$ meV. By comparing those two figures in \Fig{fig8}, one can see that $N_{1\downarrow}$ ($J=-0.32$ meV)$\approx N_{1\downarrow}$ ($J=0.32$ meV) at bias $V=0$, but both  $N_{2\downarrow}$ and $N_{2\uparrow}$ are distinctly different at $J=\pm0.32$ meV. Specifically, when $J=-0.32$ meV and $V=0$, $N_{2\downarrow}>N_{2\uparrow}$, which means that the single occupation of down-spin electron in dot 2 is sufficient and the weight of hole with the same spin in double occupation is small, thus the transfer of down-spin electrons from dot 1 to 2 will be blocked. With the bias positively increasing, as shown in \Fig{fig8}(a),  $N_{2\downarrow}$ gradually increases and approaches to 1.0 after $V>1.0$ mV,  as a consequence, the PSB effect is enhanced and the current decreases to a near zero value. Negatively increasing bias will produce a steady current at about   $-4.0$ nA after $V<-1$ mV as shown in the same figure.

On the other hand, when $J=0.32$ meV and $V=0$, $N_{2\downarrow}<N_{2\uparrow}$, which means that the single occupation of down-spin electron in dot 2 is small and the weight of hole with the same spin in double occupation is large,  thus the PSB of down-spin electrons from dot 1 to 2 will be lifted.  With the bias positively increasing, as shown in \Fig{fig8}(b),  $N_{2\downarrow}$ slowly increases and then stabilises at 0.9 after $V>1.0$ mV. It suggests that the weight of hole in double occupation keeps finite, which will induce considerable leakage current even at large positive $V$. Negatively increasing bias will produce a steady current at about   $-6.0$ nA after $V<-1$ mV, much larger than that at $J=-0.32$ meV [cf.~\Fig{fig8} (a) and (b)].  The reason may lie in the fact that antiferromagnetic interaction tends to form $S(1,1)$ which is in favor of the transition of $S(0,2)\rightarrow S(1,1)$.

\begin{figure}[!hbt]
\includegraphics [width=2.5in]{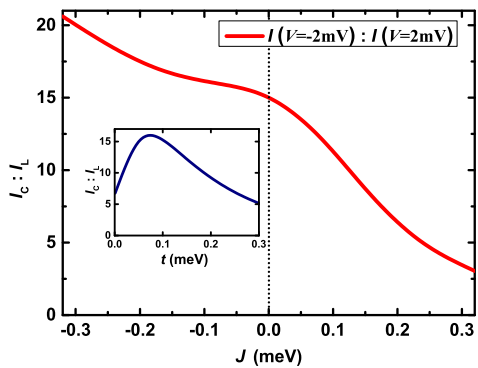}
\caption{(Color online). The dependence of the ratio $I_C(V=-2.0{\rm  mV}):I_L(V=2.0{\rm  mV})$ on $J$ at the point of $(V_1, V_2)=(-U/2, U/2)$. The insert shows the dependence of the ratio $I_C:I_L$ on $t$.}\label{fig9}
\end{figure}

We are now on the position to compare the different roles of $J$ and $t$ on PSB. For this purpose, we define a physical quantity $I_C:I_L$ as the ratio between the conductive current $I_C$ at $V=-2.0$ mV and the leakage one $I_L$ at $V=2.0$ mV at the point of $(V_1, V_2)=(-U/2, U/2)$, and then summarize the dependence of $I_C:I_L$ on $J$ in \Fig{fig9} together with $I_C:I_L$ on $t$ in its insert. In principle, the larger the ratio $I_C:I_L$ is, the smaller the leakage current, and the more distinct the PSB effect, in case that $I_C$ does not change significantly. As shown in the figure, the ratio $I_C:I_L$ can reach 21 at $J=-0.32$ meV, indicating a well-defined PSB effect, as shown in \Fig{fig6}(a) and \Fig{fig8}(a). Positively increasing $J$ will induce a continuously decreasing of $I_C:I_L$, smoothly passing through the zero point, $I_C:I_L\sim15$ at $J=0$, and finally approaching a small value $I_C:I_L\sim 3$ at $J=0.32$ meV, indicating a total lift of the PSB effect. At $J<-0.2$ meV and $J>0.2$ meV, one can see that the decrease of $I_C:I_L$ with $J$ exhibits a near-linear behavior, but a nonlinear swell around the zero point is clearly shown in the figure. It suggests that $J=0$ is a crossover point from the appearance to the lift of the PSB effect (at a fixed small $t$).

By referring the insert of \Fig{fig9}, one can see that the dependence of $I_C:I_L$ on $t$ exhibits a peak structure. More specifically,  starting from the non-PSB point at $t\sim0$ [see \Fig{fig1} and \Fig{fig3}(a)], increasing $t$ will firstly result in  a rapid increase of $I_C:I_L$ to a value about 16 at $t=0.8$, denoting a well-defined PSB effect, as shown in \Fig{fig2} and \Fig{fig3}(b). By further increasing $t$ ($t>0.8$), the $I_C:I_L$ will not increase any more, but slowly decrease to a small value about 5 at $t=0.3$. It represents that large $t$ does not mean distinct PSB effects which should only takes place at moderate $t$.  That result is consistent with the experimental measurements in literatures\cite{Hanson2007siqd,Zwanenburg2013siliconqd}.

\section{\label{sec:level3}SUMMARY}
In summarize, we systematically investigate the Pauli spin blockade in double quantum dot systems.  We start from the Anderson multiple impurity model to describe the system, fully considering the electron-electron interaction and the dot-electrode couplings. By using the hierarchical equations of motion approach, we deal with this quantum model non-perturbatively to accurately obtain the equilibrium and nonequilibrium spectral functions, occupation numbers and current, etc.

By means of a strong local magnetic field applied onto dot 1 to lift its spin degeneracy, our theory appropriately describes the physical mechanism and picture of the Pauli spin blockade effect in double quantum dot systems, followed by a general discussion without the local field applied. Then, the gate voltage manipulation of the spin blockade is elaborated in detail by our theory. Our results are proved to be qualitatively consistent with the experimental measurements.

We further extend the Anderson multiple impurity model to involve the dot-dot exchange coupling,  and carefully study its effect on the spin blockade by changing the strength and the sign of the coupling.  It is found that the ferromagnetic exchange interaction tends to enhance the spin blockade, while the antiferromagnetic one to suppress it. What is more, the Pauli spin blockade effect may be lifted by the strong antiferromagnetic exchange coupling.

\section{\label{sec:level3}ACKNOWLEDGEMENT}

The support from the NSF of China (No.11374363) and the Research Funds of Renmin University of China
(Grant No. 11XNJ026) is gratefully appreciated.

\bibliographystyle{unsrt}

\end{document}